\documentclass[10pt, final, conference, letterpaper, onecolumn, oneside]{./IEEEtran}

\usepackage{cite}
\usepackage{graphicx}
\usepackage{subfigure}
\usepackage{amsmath}
\usepackage{amssymb}
\usepackage{bm}
\usepackage{epsfig}
\usepackage{times}
\usepackage{color}
\usepackage{url}
\usepackage{array}

\newtheorem{pavikl}{\textbf{Lemma}}

\newcommand{\argmax}{\operatornamewithlimits{argmax}}

\begin{document}

\title{Analog Network Coding in General SNR Regime: Performance of Network Simplification}%
\author{\IEEEauthorblockN{Samar Agnihotri, Sidharth Jaggi, and Minghua Chen}\\%
\IEEEauthorblockA{Department of Information Engineering, The Chinese University of Hong Kong, Hong Kong\\}%
Email: samar.agnihotri@gmail.com, \{jaggi, minghua\}@ie.cuhk.edu.hk%
}

\maketitle

\begin{abstract}
We consider a communication scenario where a source communicates with a destination over a directed layered relay network. Each relay performs analog network coding where it scales and forwards the signals received at its input. In this scenario, we address the question: What portion of the maximum end-to-end achievable rate can be maintained if only a fraction of relay nodes available at each layer are used?

We consider, in particular, the Gaussian diamond network (layered network with a single layer of relay nodes) and a class of symmetric layered networks. For these networks we show that each relay layer increases the additive gap between the optimal analog network coding performance with and without network simplification (using $k$ instead of $N$ relays in each layer, $k < N$) by no more than $\log(N/k)^2$ bits and the corresponding multiplicative gap by no more than a factor of $(N/k)^2$, asymptotically (in source power). To the best of our knowledge, this work offers the first characterization of the performance of network simplification in general layered amplify-and-forward relay networks. Further, unlike most of the current approximation results that attempt to bound optimal rates either within an additive gap or a multiplicative gap, our results suggest a new rate approximation scheme that allows for the simultaneous computation of additive and multiplicative gaps.
\end{abstract}

\section{Introduction}
\label{sec:intro}
Analog network coding (ANC) extends to multihop wireless networks the idea of linear network coding \cite{103liYeungCai}, where an intermediate node sends out a linear combination of its incoming packets. In a wireless network, signals transmitted simultaneously by multiple sources add in the air. Thus, each node receives at its input a \textit{noisy sum} of these signals, \textit{i.e.} a linear combination of the received signals and noise. A communication scheme wherein each relay node merely amplifies and forwards this noisy sum is referred to as analog network coding \cite{107kattiGollakottaKatabi, 110maricGoldsmithMedard}.

The rates achievable with ANC in layered relay networks is analyzed in \cite{110maricGoldsmithMedard, 111liuCai}. In \cite{110maricGoldsmithMedard}, the achievable rate is computed under two assumptions: (A) each relay node scales the received signal to the maximum level possible subject to its transmit power constraint, (B) the nodes in all $L$ layers operate in the high-SNR regime, where $\min_{k \in l} P_{R,k} \ge 1/\delta, l = 1, \ldots, L$ for $\delta \ge 0$ and $P_{R,k}$ is the received signal power at the $k^\textrm{th}$ node. It is shown that the rate achieved under these two assumptions asymptotically (in source power) approaches network capacity. The authors in \cite{111liuCai} extend this result to the scenarios where the nodes in at most one layer do not satisfy these assumptions and show that achievable rates in such scenarios still approach the network capacity as the source power increases.

However, requiring each relay node to amplify its received signal to the upper bound of its transmit power constraint results in suboptimal performance of analog network coding in general, as we show in \cite{111agnihotriJaggiChen, 112agnihotriJaggiChen, 112agnihotriJaggiChen2}. Further, even in low-SNR regimes amplify-and-forward relaying can be capacity-achieving relay strategy in some communication scenarios, \cite{107gomadamJafar}. Therefore, it is desirable to analyze the performance of analog network coding in general layered networks, without the above two assumptions on input signal scaling factors and received SNRs. 

Analyzing the performance of analog network coding without such assumptions however, results in a computationally intractable problem in general \cite{111liuCai, 111agnihotriJaggiChen}. Therefore, to compute the maximum achievable ANC rate in general communication scenarios, in \cite{112agnihotriJaggiChen} we establish a result that significantly reduces the computational complexity of this problem. Further in \cite{112agnihotriJaggiChen2}, we propose a greedy scheme to bound from below the optimal rate achievable with analog network coding in general layered networks. For general layered relay networks, these two results allow approximating the optimal ANC rates within a constant gap from the cut-set upper bound asymptotically in the source power. These results not only allow us to exactly characterize the optimal ANC rate in a class of symmetric layered networks that cannot be addressed using existing approaches under the assumptions A and B, but allow us to do so in a computationally efficient manner.

In this paper we introduce another approach to reduce the computational complexity of approximating the maximum achievable ANC rate in general layered networks. The proposed approach is based on the notion of \textit{network simplification}, introduced in \cite{111nazarogluOzgurFragouli} to characterize fraction of the capacity of the Gaussian $N$-relay diamond network when only $k$ out of $N$ available relay nodes are used. Previously, in \textit{cooperative communication} literature as in \cite{106bletasKhistiReedLippman, 107zhaoAdveLim, 108caiShenMarkAlfa}, the notion of network simplification has been used in the amplify-and-forward relay networks to characterize the achievable rates. However, such prior work used network simplification in a restricted sense (selecting the \textit{best} single relay node among $N$ relays) and considered simple network topologies (the source communicating with the destination via a single relay node). In contrast, we provide the optimal ANC rate characterization in much general communication scenarios where any number $k$ of relay nodes in a layer are used ($k \ge 1$) and the source communicates with the destination over any number $L$ of layers of relay nodes ($L \ge 1$). In this sense, to the best of our knowledge, this is the first work to characterize the performance of network simplification in such general layered amplify-and-forward relay networks.

We show that in the Gaussian $N$-relay diamond network (layered network with a single layer of relay nodes $L=1$) and a class of symmetric layered networks, the network simplification allows us to maintain achievable rates within small additive and multiplicative gaps of the maximum ANC rate achievable when all relays in a layer are used. Along with our previous results in \cite{112agnihotriJaggiChen, 112agnihotriJaggiChen2}, where we establish that analog network coding is a capacity achieving strategy in some communication scenarios, the results in this paper establish that in those scenarios using fewer nodes in each layer, we can still maintain achievable rates within small additive and multiplicative gaps to the capacity. Further our results indicate that in general layered networks, the maximum achievable ANC rate can be tightly approximated with much lower computational complexity than with existing schemes.

\textit{Organization:} In Section~\ref{sec:sysModel} we introduce a general wireless layered relay network model and formulate the problem of maximum rate achievable with ANC in such a network. Section~\ref{sec:diamond} addresses the performance of network simplification in the Gaussian $N$-relay diamond network and computes additive and multiplicative gaps between the maximum ANC rates achievable when $N$ and $k (k < N)$, relays are used. In Section~\ref{sec:genNet} we consider a class of symmetric layered networks and compute additive and multiplicative gaps between the optimal ANC rates obtained when all relays in every layer are used and when only a fraction of such relays are used. Section~\ref{sec:conclFW} concludes the paper.

\section{System Model}
\label{sec:sysModel}
\begin{figure}[!t]
\centering
\includegraphics[width=3.0in]{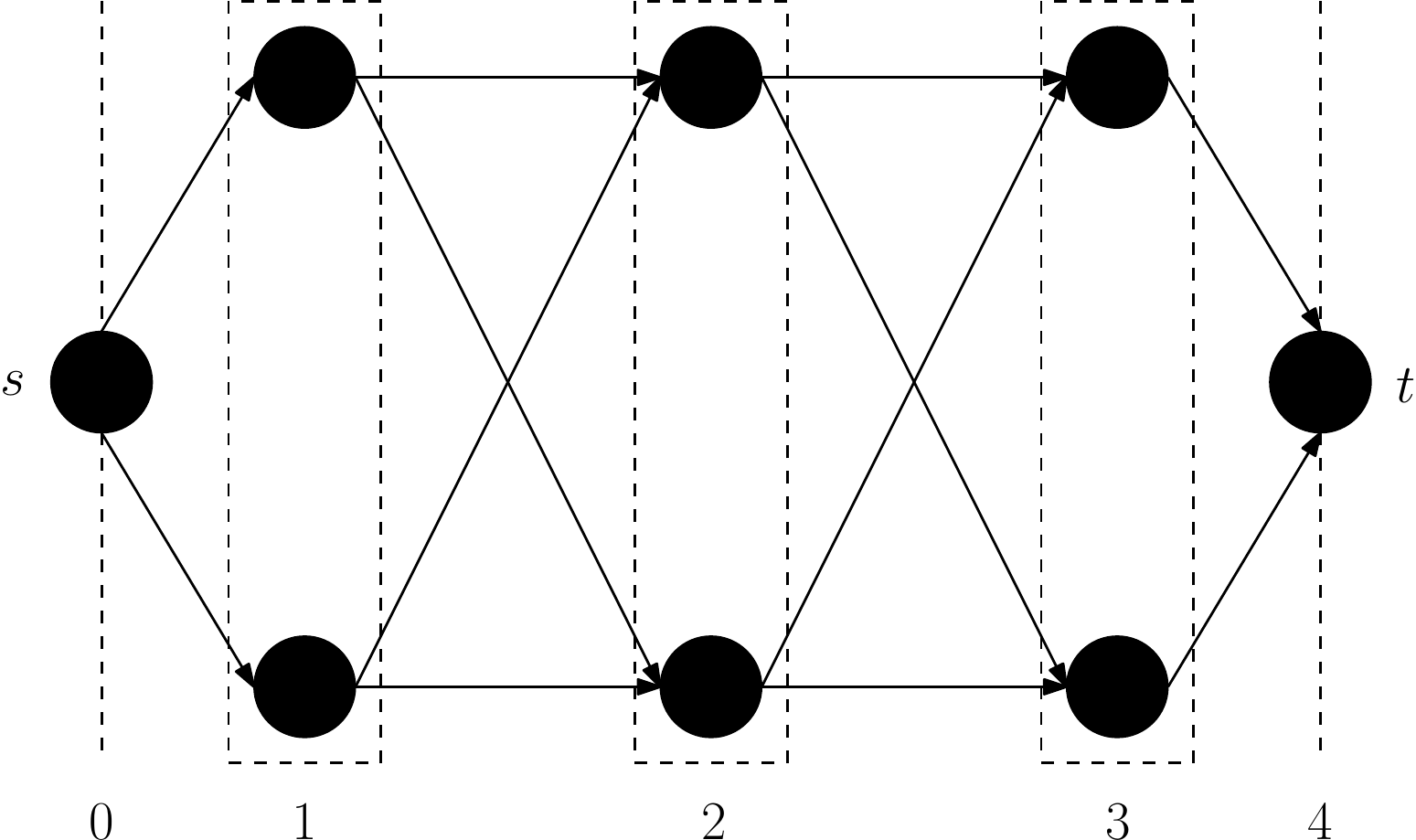}
\caption{Layered network with 3 relay layers between the source `s' and destination `t'. Each layer contains two relay nodes.}
\vspace{-0.2in}
\label{fig:layrdNetExa}
\end{figure}

Consider a $(L+2)$-layer wireless network with directed links. The source $s$ is at the layer `$0$', the destination $t$ is at the layer `$L+1$', and the relay nodes from the set $R$ are arranged in $L$ layers between them. The $l^\textrm{th}$ layer contains $n_l$ relay nodes, $\sum_{l=1}^L n_l = M$. An instance of such a network is given in Figure~\ref{fig:layrdNetExa}. Each node is assumed to have a single antenna and operate in full-duplex mode.

At instant $n$, the channel output at node $i, i \in R \cup \{t\}$, is
\begin{equation}
\label{eqn:channelOut}
y_i[n] = \sum_{j \in {\mathcal N}(i)} h_{ji} x_j[n] + z_i[n], \quad - \infty < n < \infty,
\end{equation}
where $x_j[n]$ is the channel input of the node $j$ in the neighbor set ${\mathcal N}(i)$ of node $i$. In \eqref{eqn:channelOut}, $h_{ji}$ is a real number representing the channel gain along the link from node $j$ to node $i$. It is assumed to be fixed (for example, as in a single realization of a fading process) and known throughout the network. The source symbols $x_s[n], - \infty < n < \infty$, are i.i.d. Gaussian random variables with zero mean and variance $P_s$ that satisfy an average source power constraint, $x_s[n] \sim {\cal N}(0, P_s)$. Further, $\{z_i[n]\}$ is a sequence (in $n$) of i.i.d. Gaussian random variables with $z_i[n] \sim {\cal N}(0, \sigma^2)$. We also assume that $z_i$ are independent of the input signal and of each other. We assume that the $i^{\textrm{th}}$ relay's transmit power is constrained as:
\begin{equation}
\label{eqn:pwrConstraint}
E[x_i^2[n]] \le P_i, \quad - \infty < n < \infty
\end{equation}

In analog network coding each relay node amplifies and forwards the noisy signal sum received at its input. More precisely, a relay node $i$ at instant $n+1$ transmits the scaled version of $y_i[n]$, its input at time instant $n$, as follows
\begin{equation}
\label{eqn:AFdef}
x_i[n+1] = \beta_i y_i[n], \quad 0 \le \beta_i^2 \le \beta_{i,max}^2 = P_i/P_{R,i},
\end{equation}
where $P_{R,i}$ is the received power at the node $i$ and choice of the scaling factor $\beta_i$ satisfies the power constraint \eqref{eqn:pwrConstraint}.

In layered networks, all copies of a source signal and a noise symbol introduced at a node traveling along different paths arrive at the destination with the same respective time delays. Therefore, the outputs of the source-destination channel are free of intersymbol interference. This simplifies the relation between input and output of the channel and allows us to omit the time-index while denoting the input and output signals.

Using \eqref{eqn:channelOut} and \eqref{eqn:AFdef}, the input-output channel between the source and destination can be written as
\begin{equation}
\label{eqn:sdchnl}
y_t = \bigg[\sum_{(i_1, \ldots, i_{L}) \in K_{L}} \hspace{-0.25in} h_{si_1}\beta_{i_1}h_{i_1 i_2} \ldots \beta_{i_{L}}h_{i_{L} t}\bigg] x_s + \sum_{l=1}^{L} \sum_{j=1}^{n_l}\bigg[\sum_{(i_l, \ldots, i_{L}) \in K_{lj,L}} \hspace{-0.25in} \beta_{lj} h_{lj, i_l} \ldots \beta_{L i_{L}}h_{L i_{L}, t}\bigg] z_{lj} + z_t,
\end{equation}
where $K_L$, is the set of $L$-tuples of node indices corresponding to all paths from the source to the destination with path delay $L$. Similarly, $K_{lj,L-l+1}$, is the set of $L-l+1$-tuples of node indices corresponding to all paths from the $j^{\textrm{th}}$ relay of $l^\textrm{th}$ layer to the destination with path delay $L-l+1$.

For all the paths between the source $s$ and the destination $t$, and all the paths between the $j^{\textrm{th}}$ relay of $l^\textrm{th}$ layer to the destination $t$ with path delay $L-l+1$, we introduce \textit{modified} channel gains, respectively, as follows
\begin{align}
h_s &= \sum_{(i_1, \ldots, i_{L}) \in K_{L}} h_{si_1}\beta_{i_1}h_{i_1 i_2} \ldots \beta_{i_{L}}h_{i_{L} t} \label{eqn:modChnlParams} \\
h_{lj} &= \sum_{(i_1, \ldots, i_{L-l+1}) \in K_{lj,L-l+1}} \beta_{lj} h_{lj, i_1} \ldots \beta_{L i_{L}}h_{L i_{L}, t} \label{eqn:modChnlParams2}
\end{align}

In terms of these modified channel gains\footnote{Modified channel gains for even a possibly exponential number of paths as in \eqref{eqn:modChnlParams} and \eqref{eqn:modChnlParams2} can be efficiently computed using line-graphs \cite{103koetterMedard}. Further, the number of such modified channel gains scales polynomially in the size of the graph being considered.}, the source-destination channel in \eqref{eqn:sdchnl} can be written as:
\begin{equation}
\label{eqn:chnlmod}
y_t = h_s x_s + \sum_{l=1}^{L} \sum_{j=1}^{n_l} h_{lj} z_{lj} + z_t 
\end{equation}

\textit{Problem Formulation:} For a given network-wide scaling vector $\bm{\beta}=(\beta_{li})_{1 \le l \le L, 1 \le i \le n_l}$, the achievable rate for the channel in \eqref{eqn:chnlmod} with i.i.d. Gaussian input is (\hspace{-0.001cm}\cite{110maricGoldsmithMedard, 111liuCai, 111agnihotriJaggiChen}):
\begin{equation}
\label{eqn:infoRateFin}
I(P_s, \bm{\beta}) = (1/2) \log\big(1 + SNR_t\big),
\end{equation}
where $SNR_t$, the signal-to-noise ratio at the destination $t$ is:
\begin{equation}
\label{eqn:snr}
SNR_t = \frac{P_s}{\sigma^2}\frac{h_s^2}{1 + \sum_{l=1}^{L} \sum_{j=1}^{n_l} h_{lj}^2}
\end{equation}

The maximum information-rate $I_{ANC}(P_s)$ achievable in a given layered network with i.i.d. Gaussian input is defined as the maximum of $I(P_s, \bm{\beta})$ over all feasible $\bm{\beta}$, subject to per relay transmit power constraint \eqref{eqn:AFdef}. In other words:
\begin{equation}
\label{eqn:maxAFrate}
I_{ANC}(P_s) \stackrel{def}{=} \max_{\bm{\beta}:0 \le \beta_{li}^2 \le \beta_{li, max}^2} I(P_s, \bm{\beta})
\end{equation}
It should be noted that $\beta_{li, max}$ (the maximum value of the scaling factor for $i^\textrm{th}$ node in the $l^\textrm{th}$ layer) depends on the scaling factors for the nodes in the previous $l-1$ layers.

Given the monotonicity of the $\log(\cdot)$ function (all logarithms in this work are assumed to be binary, except some cases where we explicitly denote the natural logarithm as $\ln(\cdot)$), we have
\begin{equation}
\label{eqn:eqProb}
\bm{\beta}_{opt} = \argmax_{\bm{\beta}:0 \le \beta_{li}^2 \le \beta_{li, max}^2}  I(P_s, \bm{\beta}) = \argmax_{\bm{\beta}:0 \le \beta_{li}^2 \le \beta_{li, max}^2}  SNR_t
\end{equation}
Therefore in the rest of the paper, we concern ourselves mostly with maximizing the received SNRs.

In \cite{112agnihotriJaggiChen}, we discussed the computational complexity of exactly solving the problem \eqref{eqn:maxAFrate} or equivalently the problem \eqref{eqn:eqProb}. Further, we also introduced a key result \cite[Lemma 2]{112agnihotriJaggiChen} that reduces the computational complexity of the problem of computing $\bm{\beta}_{opt}$ by computing it layer-by-layer as a solution of a cascade of subproblems. However, each of these subproblems itself is computationally hard for general network scenarios as it involves maximizing the ratio of \textit{posynomials} \cite{107boydkimVandenberghe, 105chiang}, which is known to be computationally intractable in general \cite{105chiang}. Therefore in this paper, based on the notion of network simplification \cite{111nazarogluOzgurFragouli}, we introduce an approach to reduce the computational complexity of solving each of these subproblems by selecting the best subset of $k_l$ relays among the set of $n_l$ relays in the $l^\textrm{th}$ layer. Thus the proposed scheme leads to an exponential reduction in the search space to solve each of these subproblems without significantly compromising on the optimal end-to-end performance, in terms of additive and multiplicative gaps between the corresponding performances, as we show below.

In the following, we discuss the performance of this network simplification based approach to approximate the maximum achievable ANC rate first, for the Gaussian $N$-relay networks and then, for a class of symmetric layered networks.

\section{Analog network coding in the Diamond Network: Performance of Network Simplification}
\label{sec:diamond}
\begin{figure}[!t]
\centering
\includegraphics[width=3.0in]{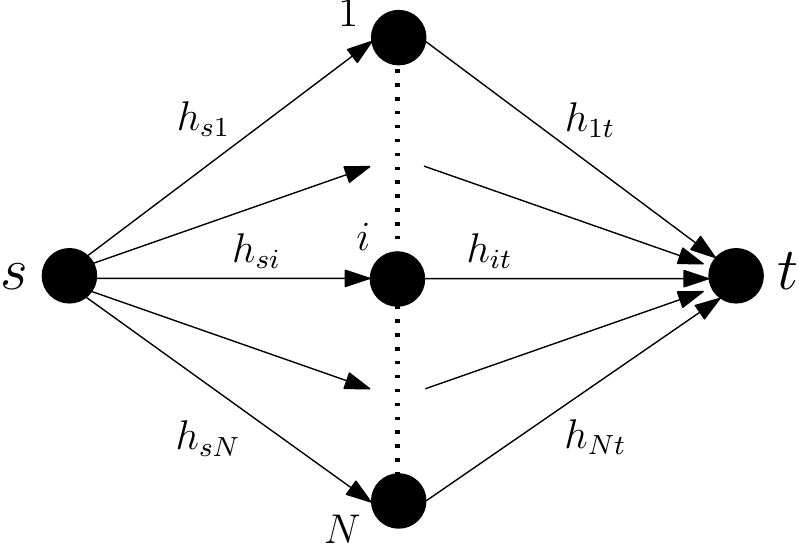}
\caption{A diamond network with $N$ relay nodes.}
\label{fig:diamond}
\vspace{-0.2in}
\end{figure}

Consider the diamond network of Figure~\ref{fig:diamond}. We can consider diamond network as a layered network with only one layer of relay nodes. Then using \eqref{eqn:modChnlParams}, \eqref{eqn:modChnlParams2}, and \eqref{eqn:snr}, we can compute the SNR at the destination $t$ for any scaling vector $\bm{\beta}$ as
\begin{equation}
\label{eqn:diamondSNR}
SNR_t = \frac{P_s}{\sigma^2} \frac{(\sum_{i=1}^N h_{si} \beta_i h_{it})^2}{1+\sum_{i=1}^N \beta_i^2 h_{it}^2}
\end{equation}

Therefore, using \eqref{eqn:maxAFrate} the problem of computing the maximum ANC rate for this network can be formulated as
\begin{equation}
\label{eqn:diamondProb}
\max_{\bm{\beta}^2 \le \bm{\beta}_{max}^2} SNR_t,
\end{equation}
where $\bm{\beta} = (\beta_1, \ldots, \beta_N)$ and $\bm{\beta}_{max} = (\beta_{1,max} \ldots, \beta_{N,max})$ with $\beta_{i,max}^2 = P_i/(h_{si}^2 P_s + \sigma^2), i \in {\mathcal N}, {\mathcal N} = \{1, \ldots, N\}$.

Equating the first-order partial derivatives of the objective function with respect to $\beta_i, i \in \mathcal{N}$, to zero, we get the following $N+1$ conditions for local extrema:
\begin{align}
&\sum_{i \in \mathcal{N}} h_{si} \beta_i h_{it} = 0 \label{eqn:b1bN} \\
&\beta_i = \frac{h_{si}/h_{it}}{\sum_{j \in \mathcal{N}\setminus\{i\}} h_{sj} \beta_j h_{jt}}\bigg(1+\sum_{j \in \mathcal{N}\setminus\{i\}} \beta_j^2 h_{it}^2\bigg) \label{eqn:bi_ito_rest}
\end{align}

In \cite{112agnihotriJaggiChen2}, we show that all points satisfying \eqref{eqn:b1bN} lead to the global minimum of the objective function in \eqref{eqn:diamondProb}. Further, we show that no real solution of the simultaneous system of equations in \eqref{eqn:bi_ito_rest} exists. In other words, no solution of \eqref{eqn:diamondProb} exists where all relay nodes transmit strictly below their respective transmit power constraints. Therefore the global maximum of the objective function occurs at one of $N$ hyperplanes (of dimension $N-1$) defined by $\beta_u = \beta_{u,max}, u \in \mathcal{N}$. Denote the hyperplane at which the objective $SNR_t$ attains its maximum as $\beta_{u^\star} = \beta_{u^\star,max}$. The details of how the hyperplane $\beta_{u^\star}$ is identified among $N$ possible candidate hyperplanes are provided in \cite{112agnihotriJaggiChen2}.

Solving the system of simultaneous equations in \eqref{eqn:bi_ito_rest} on the $(N-1)$-dimensional hyperplane defined by $\beta_{u^\star} = \beta_{u^\star,max}$ results in
\begin{equation}
\label{eqn:optBeta_on_uth_plane}
\beta_{i}^{u^\star} = \frac{h_{si}}{h_{it}} \frac{1+\beta_{u^\star,max}^2 h_{u^\star t}^2}{h_{s u^\star} \beta_{u^\star,max} h_{u^\star t}}, i \in \mathcal{N} \setminus \{u^\star\}
\end{equation}
Note that $\{\beta_{i}^{u^\star}\}$ is the set of scaling-factors for the nodes in the set of relay nodes $\mathcal{N} \setminus \{u^\star\}$ that maximizes $SNR_t$ on the hyperplane $\beta_{u^\star} = \beta_{u^\star,max}$.

Denote the scaling vector $\bm{\beta}$ at which $SNR_t$ attains its maximum on the hyperplane $\beta_{u^\star} = \beta_{u^\star,max}$ as $\bm{\beta}^{u^\star} = (\beta_{1}^{u^\star}, \ldots, \beta_{N}^{u^\star})$, where $\beta_{i}^{u^\star}$ is given by \eqref{eqn:optBeta_on_uth_plane} when $i \neq u^\star$ and $\beta_{i}^{u^\star} = \beta_{u^\star,max}$ when $i=u^\star$. Computing $SNR_t(\bm{\beta}^{u^\star})$ by substituting $\bm{\beta}^{u^\star}$ in \eqref{eqn:diamondSNR} results in
\begin{align}
SNR_t(\bm{\beta}^{u^\star}) &= (P_s/\sigma^2) \bigg[\sum_{i \in \mathcal{N}} h_{si}^2 - h_{sk}^2/(1+\beta_{k,max}^2 h_{kt}^2)\bigg] \label{eqn:unconstrainedSNR} \\
                            &< (P_s/\sigma^2)\sum_{i \in \mathcal{N}} h_{si}^2 \le (P_s/\sigma^2)N h_{si,max}^2, \label{eqn:unconstrainedSNRuprBdd}
\end{align}
where $h_{si,max} = \max_{i \in \mathcal{N}} h_{si}$.

Note that $SNR_t(\bm{\beta}^{u^\star})$ in \eqref{eqn:unconstrainedSNR} is computed when $\beta_{i}^{u^\star}$ in \eqref{eqn:optBeta_on_uth_plane} is not constrained by its corresponding upper-bound $\beta_{i,max}$. Therefore, $SNR_t(\bm{\beta}^{u^\star})$ may not be achievable in the actual system. The optimal achievable destination SNR with $N$ relays $SNR_{t,N}^{opt}$ is obtained by evaluating \eqref{eqn:diamondSNR} at $\bm{\beta}_{opt} = (\beta_{1}^{opt}, \ldots, \beta_{N}^{opt})$, where each $\beta_{i}^{opt}$ satisfies its corresponding upper bound. Theorem~1 and Lemma~2 in \cite{112agnihotriJaggiChen2} provide the exact characterization of $\bm{\beta}_{opt}$ and detail the procedure of obtaining it from $\bm{\beta}^{u^\star}$. However, $SNR_t(\bm{\beta}^{u^\star})$ provides an upper bound on $SNR_{t,N}^{opt}$, that is:
\begin{equation}
\label{eqn:optSNRuprBdd}
SNR_{t,N}^{opt} \le SNR_t(\bm{\beta}^{u^\star})
\end{equation}
However, this upper bound on $SNR_{t,N}^{opt}$ can be tight or arbitrarily loose depending on the system parameters. In particular, as the source power $P_s$ increases, this upper-bound becomes looser; on the other hand, as the source power decreases, this upper bound becomes tighter.

In \cite{111niesenDiggavi} it is observed that at high rate (or high SNR in the present setting) we should be able to approximate the maximum achievable rate within an additive gap and at low rate (or low SNR) we should be able to approximate the maximum achievable rate within a multiplicative gap. In this paper, we are particularly concerned with the variation of the destination SNR with the source power. Therefore for the ease of presentation, we compute the additive gap for large $P_s$ (high-SNR regime) and the multiplicative gap for small $P_s$ (low-SNR regime).

Let $R_N = (1/2) \log(1+SNR_{t,N}^{opt})$ denote the optimal end-to-end ANC rate achieved by using all $N$ relays in the diamond network. Similarly, let $R_k = (1/2) \log(1+SNR_{t,k}^{opt})$ denote the optimal end-to-end ANC rate achieved by using any $k$ relays out of available $N$ relays in the diamond network. In the following, we compute an upper bounds on additive gap ($R_N - R_k$) and multiplicative gap ($R_N/R_k$) for large and small $P_s$, respectively.

\subsection{An Upper Bound on the Additive Gap}
\label{subsec:diamondAddGap}
It follows from the definition of $\beta_i$, the scaling-factor for the $i^\textrm{th}$ node, that $\beta_i^2 \sim \frac{1}{P_s}$. Then, \eqref{eqn:optBeta_on_uth_plane} implies that for a \textit{sufficiently large} $P_s$, say $P_s^l$, every $\beta_{i}^{u^\star}$ violates its corresponding upper-bound and in an optimal scaling-vector $\bm{\beta}_{opt}$, we have $\beta_{i}^{opt}=\beta_{i,max}$ as $P_s > P_s^l$. Therefore, with $N$ relays, we have for the optimal SNR at the destination:
\begin{align}
SNR_{t,N}^{opt} &= \frac{P_s}{\sigma^2} \frac{(\sum_{i \in {\mathcal N}} h_{si} \beta_{i,max} h_{it})^2}{1+\sum_{i \in {\mathcal N}} \beta_{i,max}^2 h_{it}^2} \nonumber \\
                &\le \frac{P_s}{\sigma^2} \frac{N^2(h_{si,max} \beta_{i,max}^{max} h_{it,max})^2}{1+ N (\beta_{i,max}^{min} h_{it,min})^2}, \label{eqn:optNSNRuprBdd}
\end{align}
where $h_{si,max} = \max_{i \in \mathcal{N}} h_{si}$, $h_{it,min} = \min_{i \in \mathcal{N}} h_{it}$, $h_{it,max} = \max_{i \in \mathcal{N}} h_{it}$, $\beta_{i,max}^{min} = \min_{i \in \mathcal{N}} \beta_{i,max}$, and $\beta_{i,max}^{max} = \max_{i \in \mathcal{N}} \beta_{i,max}$.

Consider $SNR_{t,k}^{opt}$, the optimal achievable SNR at the destination when any subset $\mathcal K$ of $k$ relays is chosen from the set $\mathcal N$ of all available relays. In other words, $SNR_{t,k}^{opt}$ is the SNR expression in \eqref{eqn:diamondSNR} evaluated at $\bm{\beta}_{opt}$ corresponding to subset $\mathcal K$.
Then, we have:
\begin{align}
SNR_{t,k}^{opt} &= \frac{P_s}{\sigma^2} \frac{(\sum_{i \in {\mathcal K}} h_{si} \beta_{i,max} h_{it})^2}{1+\sum_{i \in {\mathcal K}} \beta_{i,max}^2 h_{it}^2} \nonumber \\
                &\ge \frac{P_s}{\sigma^2} \frac{k^2(h_{si,min} \beta_{i,max}^{min} h_{it,min})^2}{1+ k (\beta_{i,max}^{max} h_{it,max})^2 }, \label{eqn:optkSNRlwrBdd1}
\end{align}
where $h_{si,min} = \min_{i \in \mathcal{N}} h_{si}$.

Therefore, from \eqref{eqn:optNSNRuprBdd} and \eqref{eqn:optkSNRlwrBdd1}, we have
\begin{align}
\frac{SNR_{t,N}^{opt}}{SNR_{t,k}^{opt}} &\le \bigg(\frac{N}{k}\bigg)^2 \frac{(h_{si,max} \beta_{i,max}^{max} h_{it,max})^2}{(h_{si,min} \beta_{i,max}^{min} h_{it,min})^2} \frac{1+ k (\beta_{i,max}^{max} h_{it,max})^2}{1+ N (\beta_{i,max}^{min} h_{it,min})^2} \nonumber \\
                                        &\overset{(a)}{\le} \bigg(\frac{N}{k}\bigg)^2 \alpha_1, \quad \alpha_1 = \frac{(h_{si,max}^2 h_{it,max})^2}{(h_{si,min}^2 h_{it,min})^2}, \label{eqn:diamondAddSNRratio}
\end{align}
where $(a)$ follows from $\frac{(\beta_{i,max}^{max})^2}{(\beta_{i,max})^2} \le \frac{h_{si,max}^2}{h_{si,min}^2}$ and $\frac{1+ k (\beta_{i,max}^{max} h_{it,max})^2}{1+ N (\beta_{i,max}^{min} h_{it,min})^2} \le 1$. Note that parameter $\alpha_1$ characterizes the asymmetry in the network and in general, $\alpha_1 \ge 1$.

Using \eqref{eqn:diamondAddSNRratio}, we compute the upper-bound on additive gap, $R_N - R_K$, with two approaches: direct and indirect.

\noindent\textbf{\textit{Direct Approach:}} We compute $R_N - R_K$ as follows:
\begin{align}
R_N - R_k &= \frac{1}{2} \log \frac{1+SNR_{t,N}^{opt}}{1+SNR_{t,k}^{opt}} \nonumber \\
          &\le \frac{1}{2} \log \frac{1+(\frac{N}{k})^2 \alpha_1 SNR_{t,k}^{opt}}{1+SNR_{t,k}^{opt}} \nonumber \\
          &\le \frac{1}{2} \log ((N/k)^2 \alpha_1) \nonumber \\
          &=  \frac{1}{2\ln 2} (2\ln (N/k) + \ln \alpha_1) \label{eqn:diamondDirectAddGap}
\end{align}

\noindent\textbf{\textit{Indirect Approach:}} We compute an upper-bound on $R_N - R_K$ by first expressing it as
\begin{equation*}
R_N - R_K = (R_N - R_{N-1}) + \ldots + (R_{k+1} - R_k)
\end{equation*}
and then summing up the upper-bound on each of these partial gaps. Consider $(R_{k+1} - R_k)$:
\begin{align}
R_{k+1} - R_k &\le \frac{1}{2} \log \frac{1+(1+1/k)^2 \alpha_1 SNR_{t,k}^{opt}}{1+SNR_{t,k}^{opt}} \nonumber \\
              &\le \frac{1}{2} \log\bigg[\bigg(1+\frac{1}{k}\bigg)^2 \alpha_1\bigg] \nonumber \\
              &\lessapprox \frac{1}{2 \ln 2} (\frac{2}{k} + \ln \alpha_1)
\end{align}
This leads to
\begin{align}
R_N - R_K &\le \frac{1}{2 \ln 2} \big(\sum_{i=k}^{N-1} \frac{2}{i} + (N-k) \ln \alpha_1\big) \nonumber \\
          &\le \frac{1}{2 \ln 2}\big[2(H_{N-1} - H_{k-1}) + (N-k) \ln \alpha_1 \big], \label{eqn:diamondIndirectAddGap}
\end{align}
where $H_n$ is the Harmonic number, $\lim_{n \rightarrow \infty} H_n \sim \ln n + \gamma$ and $\gamma$ is \textit{Euler-Mascheroni} constant \cite{097knuth}.

Combining the upper-bounds on $R_N - R_k$ computed in \eqref{eqn:diamondDirectAddGap} and \eqref{eqn:diamondIndirectAddGap}, we have the following upper-bound on the additive gap for \textit{sufficiently large} $P_s$:
\begin{equation}
\label{eqn:diamondAddGap}
R_N - R_k \le \frac{1}{2\ln 2} \min\{2\ln (N/k) + \ln \alpha_1, 2(H_{N-1} - H_{k-1}) + (N-k) \ln \alpha_1\}
\end{equation}

\subsection{An Upper Bound on the Multiplicative Gap}
\label{subsec:diamondMultGap}
Equation \eqref{eqn:optBeta_on_uth_plane} implies that for a \textit{sufficiently small} $P_s$, say $P_s^s$, each $\beta_{i}^{u^\star}$ is within its corresponding upper-bound. Therefore, the upper bound on $SNR_{t,N}^{opt}$ in \eqref{eqn:optSNRuprBdd} is tight.

Let $SNR_{t,k}^{max}$ denote the destination SNR when each relay in the subset $\mathcal K$ transmits at the upper-bound of its transmit power constraint, {\it i.e.} $\beta_i = \beta_{i,max}, i \in {\mathcal K}$. However, all nodes transmitting at the upper-bound of their transmit power constraint results in suboptimal performance, as we showed in \cite{111agnihotriJaggiChen, 112agnihotriJaggiChen}. Therefore, we have
\begin{align}
SNR_{t,k}^{opt} \ge SNR_{t,k}^{max} &= \frac{P_s}{\sigma^2} \frac{(\sum_{i \in {\mathcal K}} h_{si} \beta_{i,max} h_{it})^2}{1+\sum_{i \in {\mathcal K}} \beta_{i,max}^2 h_{it}^2} \nonumber \\
                                    &\ge \frac{P_s}{\sigma^2} \frac{k^2(h_{si,min} \beta_{i,max}^{min} h_{it,min})^2}{1+ k (\beta_{i,max}^{max} h_{it,max})^2 } \label{eqn:optkSNRlwrBdd2}
\end{align}

Therefore, from \eqref{eqn:unconstrainedSNRuprBdd}, \eqref{eqn:optSNRuprBdd}, and \eqref{eqn:optkSNRlwrBdd2}, we have
\begin{align}
\frac{SNR_{t,N}^{opt}}{SNR_{t,k}^{opt}} &\le N h_{si,max}^2 \frac{1+ k (\beta_{i,max}^{max} h_{it,max})^2}{k^2(h_{si,min} \beta_{i,max}^{min} h_{it,min})^2} \nonumber \\
                                        &= \frac{N}{k} \frac{(h_{si,max} \beta_{i,max}^{max} h_{it,max})^2}{(h_{si,min} \beta_{i,max}^{min} h_{it,min})^2} \bigg(1+ \frac{1}{k (\beta_{i,max}^{max} h_{it,max})^2}\bigg), \nonumber \\
                                        &\le \frac{N}{k} \frac{(h_{si,max} h_{it,max})^2}{(h_{si,min} h_{it,min})^2} \frac{h_{si,max}^2 + \sigma^2}{h_{si,min}^2 + \sigma^2} \bigg(1+ \frac{1}{k (\beta_{i,max}^{max} h_{it,max})^2}\bigg), \nonumber \\
                                        &\le \frac{N}{k} \alpha_2 \bigg(1+ \frac{\gamma}{k}\bigg), \alpha_2 = \frac{(h_{si,max} h_{it,max})^2}{(h_{si,min} h_{it,min})^2} \frac{h_{si,max}^2 + \sigma^2}{h_{si,min}^2 + \sigma^2}, \gamma =  \frac{1}{(\beta_{i,max}^{max} h_{it,max})^2}, \label{eqn:diamondMultSNRratio}
\end{align}
where parameter $\alpha_2$ characterizes the asymmetry in the network and in general, $\alpha_2 \ge 1$.

Using \eqref{eqn:diamondMultSNRratio}, we compute the upper-bound on multiplicative gap, $R_N/R_k$, as follows:
\begin{align}
\frac{R_N}{R_k} &= \frac{\log(1+SNR_{t,N}^{opt})}{\log(1+SNR_{t,k}^{opt})} \nonumber \\
                &\le \frac{\log(1+\frac{N}{k} \alpha_2 (1+ \gamma/k)SNR_{t,k}^{opt})}{\log(1+SNR_{t,k}^{opt})} \nonumber \\
                &\stackrel{(a)}{\lessapprox} 1 + \bigg[\frac{N}{k} \alpha_2 \bigg(1+ \frac{\gamma}{k}\bigg) - 1 \bigg] \frac{SNR_{t,k}^{opt}}{(1+SNR_{t,k}^{opt})\ln(1+SNR_{t,k}^{opt})} \nonumber \\
                &\stackrel{(b)}{\le} \frac{N}{k} \alpha_2 \bigg(1+ \frac{\gamma}{k}\bigg) \le \frac{N}{k} \alpha_2 (1 + \gamma), \label{eqn:diamondMultGap}
\end{align}
where $(a)$ follow when $SNR_{t,k}^{opt} \rightarrow 0$ and $(b)$ follows from noting that $\frac{z}{(1+z)\log(1+z)} \le 1$ for all $z \ge 0$.

The results of the previous two subsections are collected in the following lemma that is our first main result.

\begin{pavikl}
\label{lemma:diamondGaps}
For the Gaussian $N$-relay diamond network, the additive and multiplicative gaps between the optimal performance of analog network coding obtained with and without network simplification are bounded from above, respectively, as
\begin{align*}
R_N - R_k &\le \frac{1}{2\ln 2} \min\{2\ln (N/k) + \ln \alpha_1, 2(H_{N-1} - H_{k-1}) + (N-k) \ln \alpha_1\} \\
\frac{R_N}{R_k} &\le \frac{N}{k} \alpha_2 \bigg(1+ \frac{\gamma}{k}\bigg),
\end{align*}
where $\alpha_1$ and $\alpha_2$ are two parameters characterizing the asymmetry in the network.
\end{pavikl}

\textbf{\textit{Example 1:}} Consider the Gaussian $N$-relay diamond network in the symmetric configuration, in which the channel gains to the relays are equal ($h_{si} = h, i \in \mathcal{N}$) and the channel gains from the relays are equal ($h_{it} = g, i \in \mathcal{N}$) and in general $h \neq g$. Also assume that the transmit power constraint on each relay is the same, {\it i.e.} $E[x_i^2] \le P$. Note that in this setting $\beta_{i,max}^2 = \beta^2 = P/(h^2 P_s + \sigma^2)$ and $\gamma = 1/(\beta g)^2$. Also, the parameters $\alpha_1$ and $\alpha_2$ that characterize the asymmetry in the network are $\alpha_1 = \alpha_2 = 1$. Therefore in this setting, using \eqref{eqn:diamondAddGap} and \eqref{eqn:diamondMultGap} we obtain the following upper bounds on additive and multiplicative gaps between $R_N$ and $R_k$ when $x \ge 1$:
\begin{align*}
R_N - R_k &\le \frac{1}{\ln 2}(H_{N-1} - H_{k-1}) \stackrel{(a)}{\sim} \log \frac{N}{k}\\
\frac{R_N}{R_k} &\le \frac{N}{k} (1+ \gamma/k) \le \frac{N}{k}(1 + \gamma),
\end{align*}
where $(a)$ holds asymptotically in $N$ and $k$. It is interesting to note that these bounds exactly coincide with the similar bounds obtained from the first principles for this system configuration, as an interested reader can verify. {\hspace*{\fill}~\IEEEQEDclosed\par}

\section{Performance of Network Simplification in General Layered Networks}
\label{sec:genNet}
\begin{figure}[!t]
\centering
\includegraphics[width=3.5in]{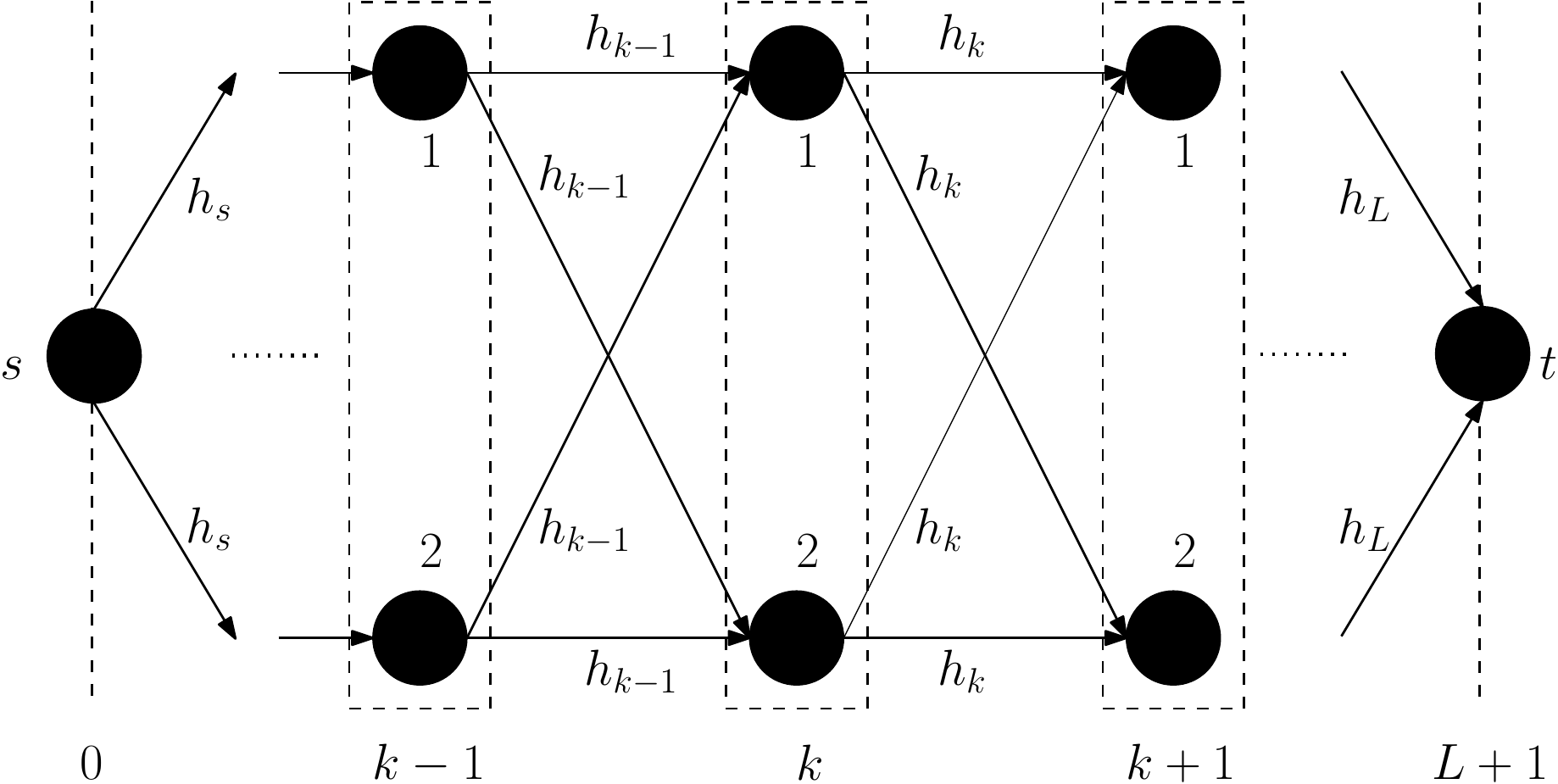}
\caption{An ECGAL network of $L+2$ layers, with the source $s$ in layer `$0$', the destination $t$ in layer `$L+1$', and $L$ layers consisting of two relay nodes each between them. The channel gains along all links between two adjacent layers are equal.}
\vspace{-0.2in}
\label{fig:ECGALnet}
\end{figure}

In this section we analyze the performance of network simplification in a class of symmetric layered networks with one or more layers of relays between the source and destination and each relay performing analog network coding on its input signal. The particular class of symmetric networks we consider here are defined such that channel gains along all links between the nodes in two adjacent layers are equal. This is a generalization of the symmetric diamond network configuration considered in Example~1 above. We introduced such networks in \cite{112agnihotriJaggiChen} and called them \textit{``Equal Channel Gains between Adjacent Layers (ECGAL)''} networks. Figure~\ref{fig:ECGALnet} provides an illustration of such networks.

For the ease of presentation, consider ECGAL networks where each layer of relay nodes has $N$ relays and all relay nodes have the same transmit power constraint $EX^2 \le P$.

Using the following result we establish in \cite{112agnihotriJaggiChen}, we can obtain the optimal solution $\bm{\beta}_{opt}^N$ of problem \eqref{eqn:eqProb} for an ECGAL network with $N$ nodes in each relay layer.

\begin{pavikl}[Lemma 2 in \cite{112agnihotriJaggiChen}]
Consider a layered relay network of $L+2$ layers, with the source $s$ in layer `$0$', the destination $t$ in layer `$L+1$', and $L$ layers of relay nodes between them. The $l^\textrm{th}$ layer contains $n_l$ nodes, $n_0 = n_{L+1} = 1$. A network-wide scaling vector $\bm{\beta}_{opt}=(\bm{\beta}_{1}^{opt}, \ldots, \bm{\beta}_{L}^{opt})$ that solves \eqref{eqn:eqProb} for this network, can be computed recursively for $1 \le l \le L$ as
\begin{equation*}
\bm{\beta}_{l}^{opt} = \argmax_{\bm{\beta}_{l}^2 \le \bm{\beta}_{l,max}^2} \prod_{i=1}^{n_{l+1}}(1+SNR_{l+1, i}(\bm{\beta}_{1}^{opt}, \ldots, \bm{\beta}_{l-1}^{opt}, \bm{\beta}_{l})),
\end{equation*}
where $\bm{\beta}_{l}^{opt}$ is the subvector of optimal scaling factors for the nodes in the $l^\textrm{th}$ layer, $\bm{\beta}_{l}^{opt} = (\beta_{l 1}^{opt}, \ldots, \beta_{l n_l}^{opt})$ and constraints $\bm{\beta}_{l}^2 \le \bm{\beta}_{l,max}^2$ are component-wise $\beta_{l i}^2 \le \beta_{l i,max}^2$.   {\hspace*{\fill}~\IEEEQEDclosed\par}
\end{pavikl}

The solution $\bm{\beta}_{opt}^N$ has the property that under it all relays in a layer use the same scaling factor and it is equal to the maximum value of the scaling factor for the nodes in the layer, {\it i.e.} for each layer $l, 1 \le l \le L$, and every node $i$ in the $l^\textrm{th}$ layer, $1 \le i \le N$, $(\beta_{li}^N)^2 = (\beta_{l}^N)^2 = (\beta_{l,max}^N)^2$, where
\begin{equation}
\label{eqn:optBetal4ecgalN}
(\beta_{l,max}^N)^2 = (\beta_{l}^N)^2 = \frac{P/\sigma^2}{\big[h_s \prod\limits_{i=1}^{l-1}(N \beta_i h_i)\big]^2 \frac{P_s}{\sigma^2} + N \sum\limits_{i=1}^{l-1} (\beta_i h_i  \prod\limits_{j=i+1}^{l-1}(N \beta_j h_j))^2 + 1} 
\end{equation}
Therefore, $\bm{\beta}_{opt}^N$ can be written as $\bm{\beta}_{opt}^N=(\beta_{1}^N, \ldots, \beta_{L}^N)$. We use superscript $N$ to emphasize that the optimal scaling factors in \eqref{eqn:optBetal4ecgalN} are computed for $N$ nodes in each layer.

The optimal SNR at the destination with $N$ relays in each layer of a layered network with $L$ relay layers is given as follows
\begin{equation}
\label{eqn:optDestSNR4ecgalN}
SNR_{t,N}^{opt} = \frac{P_s}{\sigma^2} \frac{\bigg(h_s \prod\limits_{l=1}^{L}(N \beta_l^N h_l)\bigg)^2}{1 + N \sum\limits_{l=1}^{L} (\beta_l^N h_l \prod\limits_{j=l+1}^{L}(N \beta_j^N h_j))^2}
\end{equation}
where $\beta_l^N$ is given as in \eqref{eqn:optBetal4ecgalN}.

Now consider the network simplification scenario where only $k$ out of $N$ available relays in each layer are used. Using Lemma 2 in \cite{112agnihotriJaggiChen}, we can solve problem \eqref{eqn:eqProb} for this simplified ECGAL network and obtain the optimal solution $\bm{\beta}_{opt}^k = (\beta_{1}^k, \ldots, \beta_{L}^k)$, where $\beta_l^k$ is the optimum scaling factor for all nodes in layer $l$ and is given as follows:
\begin{equation}
\label{eqn:optBetal4ecgalk}
(\beta_{l}^k)^2 = \frac{P/\sigma^2}{\big[h_s \prod\limits_{i=1}^{l-1}(k \beta_i h_i)\big]^2 \frac{P_s}{\sigma^2} + k \sum\limits_{i=1}^{l-1} (\beta_i h_i  \prod\limits_{j=i+1}^{l-1}(k \beta_j h_j))^2 + 1} 
\end{equation}
The corresponding optimal SNR at the destination is given as:
\begin{equation}
\label{eqn:optDestSNR4ecgalk}
SNR_{t,k}^{opt} = \frac{P_s}{\sigma^2} \frac{\bigg(h_s \prod\limits_{l=1}^{L}(k \beta_l^k h_l)\bigg)^2}{1 + k \sum\limits_{l=1}^{L} (\beta_l^k h_l \prod\limits_{j=l+1}^{L}(k \beta_j^k h_j))^2}
\end{equation}

Using \eqref{eqn:optDestSNR4ecgalN} and \eqref{eqn:optDestSNR4ecgalk}, we can compute the ratio of the optimal destination SNRs obtained when all $N$ relays in every layer are used and when only $k$ out of $N$ relays in every layer are used. After a few steps of algebraic manipulation, we get
\begin{equation}
\label{eqn:ecgalSNRratio}
\frac{SNR_{t,N}^{opt}}{SNR_{t,k}^{opt}} = \bigg(\frac{N}{k} \bigg)^{2L-1} {\displaystyle \frac{\sum\limits_{l=1}^L \frac{1}{\big(\prod\limits_{j=1}^{l-1}(k \beta_j^k h_j)\big)^2} + \frac{k}{\big(\prod\limits_{l=1}^{L}(k \beta_l^k h_l)\big)^2}}{\sum\limits_{l=1}^L \frac{1}{\big(\prod\limits_{j=1}^{l-1}(N \beta_j^N h_j)\big)^2} + \frac{N}{\big(\prod\limits_{l=1}^{L}(N \beta_l^N h_l)\big)^2}}}
\end{equation}

Let $R_N = (1/2) \log(1+SNR_{t,N}^{opt})$ denote the optimal end-to-end ANC rate achieved by using all $N$ relays every layer of an ECGAL network. Similarly, let $R_k = (1/2) \log(1+SNR_{t,k}^{opt})$ denote the optimal end-to-end ANC rate achieved by using any $k$ relays out of available $N$ relays in every layer of an ECGAL network. Arguing as in Section~\ref{sec:diamond}, in the following we compute the additive gap $R_N - R_k$ for large $P_s$ and the multiplicative gap $R_N/R_k$ for small $P_s$.

Before we discuss the upper-bounds on the additive and multiplicative gaps for ECGAL networks with arbitrary number of relay layers, consider the following example where we compute such bounds for an ECGAL network with two layers of relay nodes ($L=2$) for any $N$ and $k$.

\textbf{\textit{Example 2:}} Consider an ECGAL network with two layers of relay nodes between the source and the destination, $L=2$. Using \eqref{eqn:ecgalSNRratio}, we have for this network
\begin{equation*}
\frac{SNR_{t,N}^{opt}}{SNR_{t,k}^{opt}} = \bigg(\frac{N}{k} \bigg)^{\!3} \frac{1+{\displaystyle \frac{1}{k^2 h_1^2 \beta_1^2}} + {\displaystyle \frac{1}{k^3 h_1^2 h_2^2 \beta_1^2 (\beta_2^k)^2}}}{1+{\displaystyle \frac{1}{N^2 h_1^2 \beta_1^2}} + {\displaystyle \frac{1}{N^3 h_1^2 h_2^2 \beta_1^2 (\beta_2^N)^2}}}
\end{equation*}
Substituting for $\beta_1, \beta_2^N$, and $\beta_2^k$ from \eqref{eqn:optDestSNR4ecgalN} and \eqref{eqn:optDestSNR4ecgalk} in the above expression results in
\begin{align}
&\mbox{(for } P_s \rightarrow \infty) \qquad \frac{SNR_{t,N}^{opt}}{SNR_{t,k}^{opt}} = \bigg(\frac{N}{k} \bigg)^{\!4} \frac{1+ \frac{h_2^2}{k h_1^2} + \frac{\sigma^2}{k^2 h_1^2 P}}{1+ \frac{h_2^2}{N h_1^2} + \frac{\sigma^2}{N^2 h_1^2 P}} \label{eqn:2lyrEcgalSNRratioPs2infty} \\
&\mbox{(for } P_s \rightarrow 0) \qquad \frac{SNR_{t,N}^{opt}}{SNR_{t,k}^{opt}} = \bigg(\frac{N}{k} \bigg)^{\!3} \frac{1+\frac{\sigma^2}{k^2 P}\big(\frac{1}{h_1^2}+\frac{1}{h_2^2}+\frac{\sigma^2}{k h_1^2 h_2^2 P}\big)}{1+\frac{\sigma^2}{N^2 P}\big(\frac{1}{h_1^2}+\frac{1}{h_2^2}+\frac{\sigma^2}{N h_1^2 h_2^2 P}\big)} \label{eqn:2lyrEcgalSNRratioPs2zero}
\end{align}

Following the same sequence of steps as in the direct approach to compute the upper bound on the additive and multiplicative gaps for the diamond network in subsections~\ref{subsec:diamondAddGap} and \ref{subsec:diamondMultGap}, respectively, using \eqref{eqn:2lyrEcgalSNRratioPs2infty} and \eqref{eqn:2lyrEcgalSNRratioPs2zero}, we get the following upper bounds on the corresponding additive and multiplicative gaps for the two layer ECGAL network:
\begin{align*}
R_N - R_k &\le \frac{1}{2 \ln 2} \bigg(4 \ln \frac{N}{k} + \ln \frac{1+ \frac{h_2^2}{k h_1^2} + \frac{\sigma^2}{k^2 h_1^2 P}}{1+ \frac{h_2^2}{N h_1^2} + \frac{\sigma^2}{N^2 h_1^2 P}}\bigg) \\
\frac{R_N}{R_k} &\le \bigg(\frac{N}{k} \bigg)^{\!3} \frac{1+\frac{\sigma^2}{k^2 P}\big(\frac{1}{h_1^2}+\frac{1}{h_2^2}+\frac{\sigma^2}{k h_1^2 h_2^2 P}\big)}{1+\frac{\sigma^2}{N^2 P}\big(\frac{1}{h_1^2}+\frac{1}{h_2^2}+\frac{\sigma^2}{N h_1^2 h_2^2 P}\big)}
\end{align*}
for arbitrary $N$ and $k$ relays in each layer. {\hspace*{\fill}~\IEEEQEDclosed\par}

However, for ECGAL networks with arbitrary number of relay layers, it is analytically hard to compute such upper-bounds on the additive and multiplicative gaps between the optimal end-to-end performances with and without network simplification for any $N$ and $k$. Therefore, in the following we attempt to analyze the scaling behavior of such upper bounds with large $N$ and $k$.
 
Consider the asymptotic behavior of $\frac{SNR_{t,N}^{opt}}{SNR_{t,k}^{opt}}$ with $P_s$ and large $N$, and $k$. Using \eqref{eqn:optBetal4ecgalN} and \eqref{eqn:optBetal4ecgalk}, we have for large $N$ and $k$:
\begin{align}
&\mbox{(for } P_s \rightarrow \infty) \qquad \prod_{i=1}^l (\beta_i^N)^2 = \frac{P/(h_s^2 P_s)}{N^{2(l-1)} \prod_{i=1}^{l-1} h_i^2}, \qquad \prod_{i=1}^l (\beta_i^k)^2 = \frac{P/(h_s^2 P_s)}{k^{2(l-1)} \prod_{i=1}^{l-1} h_i^2}, \quad 1 \le l \le L \label{eqn:betaProdPs2infty} \\
&\mbox{(for } P_s \rightarrow 0) \qquad \prod_{i=1}^l (\beta_i^N)^2 = \frac{P/\sigma^2}{N^{2(l-1)-1} \prod_{i=1}^{l-1} h_i^2}, \qquad \prod_{i=1}^l (\beta_i^k)^2 = \frac{P/\sigma^2}{k^{2(l-1)-1} \prod_{i=1}^{l-1} h_i^2}, \quad 1 \le l \le L \label{eqn:betaProdPs2zero}
\end{align}
Then, from \eqref{eqn:ecgalSNRratio} and \eqref{eqn:betaProdPs2infty} we obtain for large $N$ and $k$, and $P_s \rightarrow \infty$:
\begin{equation}
\label{eqn:ecgalSNRratioPs2infty}
\frac{SNR_{t,N}^{opt}}{SNR_{t,k}^{opt}} \sim \bigg(\frac{N}{k} \bigg)^{2L} \bigg[1 + a \bigg(\frac{1}{k} - \frac{1}{N}\bigg)\bigg], \quad \mbox{where } a = h_L^2 \sum_{i=1}^{L-1} \frac{1}{h_i^2}
\end{equation}
and from \eqref{eqn:ecgalSNRratio} and \eqref{eqn:betaProdPs2zero} we obtain for large $N$ and $k$, and $P_s \rightarrow 0$:
\begin{equation}
\label{eqn:ecgalSNRratioPs2zero}
\frac{SNR_{t,N}^{opt}}{SNR_{t,k}^{opt}} \sim \bigg(\frac{N}{k} \bigg)^{2L-1} \bigg[1 + b \bigg(\frac{1}{k^2} - \frac{1}{N^2}\bigg)\bigg], \quad \mbox{where } b = \frac{\sigma^2}{P} \bigg(\frac{1}{h_1^2} + \frac{1}{h_L^2}\bigg)
\end{equation}

With these results we are now ready to compute the upper bounds on additive and multiplicative gaps between the optimal performance of analog network coding with and without network simplification.

First, we consider the upper bound on the additive gap. Following the same sequence of steps as in the direct approach to compute the upper bound on the additive gap for the diamond network in subsection~\ref{subsec:diamondAddGap}, using \eqref{eqn:ecgalSNRratioPs2infty} we obtain the following upper bound on the additive gap $\overline{R}_N - \overline{R}_k$ for $P_s \rightarrow \infty$ and asymptotically large $N$ and $k$ satisfying $N, k = o(P_s)$:
\begin{equation}
\label{eqn:ecgalDirectAddGap}
\overline{R}_N - \overline{R}_k \le \frac{1}{2\ln 2} \bigg\{2L \ln \frac{N}{k} + \ln \bigg[1+a \bigg(\frac{1}{k} - \frac{1}{N}\bigg)\bigg]\bigg\}
\end{equation}

Next, we consider the upper bound on the multiplicative gap. Following the same sequence of steps as in subsection~\ref{subsec:diamondMultGap} to compute the upper bound on the multiplicative gap for the diamond network, using \eqref{eqn:ecgalSNRratioPs2zero} we obtain the following upper bound on the multiplicative gap $\overline{R}_N/\overline{R}_k$ for $P_s \rightarrow 0$ and asymptotically large $N$ and $k$ satisfying $N, k = o(1/P_s)$:
\begin{equation}
\label{eqn:ecgalDirectMultGap}
\frac{\overline{R}_N}{\overline{R}_k} \le \bigg(\frac{N}{k} \bigg)^{2L-1} \bigg[1+b \bigg(\frac{1}{k^2} - \frac{1}{N^2}\bigg)\bigg]
\end{equation}

These results on the asymptotic behavior of the additive and multiplicative gaps are collected in the following lemma that is the second main result of this paper.
\begin{pavikl}
\label{lemma:ecgalGaps}
For ECGAL network, the asymptotic additive and multiplicative gaps between the optimal performance of analog network coding obtained with and without network simplification are bounded from above as
\begin{align*}
\overline{R}_N - \overline{R}_k &\le \frac{1}{2\ln 2} \bigg\{2L \ln \frac{N}{k} + \ln \bigg[1+a \bigg(\frac{1}{k} - \frac{1}{N}\bigg)\bigg]\bigg\} \le \frac{1}{2\ln 2} \bigg[2L \ln \frac{N}{k} + \ln (1+a)\bigg], \\
\frac{\overline{R}_N}{\overline{R}_k} &\le \bigg(\frac{N}{k} \bigg)^{2L-1} \bigg[1+b \bigg(\frac{1}{k^2} - \frac{1}{N^2}\bigg)\bigg] \le \bigg(\frac{N}{k} \bigg)^{2L-1} (1+b),
\end{align*}
where $a = h_L^2 \sum_{i=1}^{L-1} 1/h_i^2$ and $b = \frac{\sigma^2}{P} (1/h_1^2 + 1/h_L^2)$ are constants depending on the system parameters.
\end{pavikl}

\textit{Remark 1:} These bounds imply that each relay layer increases the additive gap by $2\log(N/k)$ and the multiplicative gap by a factor of $(N/k)^2$. Thus the per layer contribution to the gaps is in agreement with the corresponding results obtained for the diamond network (with a single relay layer) in Section~\ref{sec:diamond}. For instance, for the network in \textit{Example 1} in Section~\ref{sec:diamond}, using the results of this section we obtain
\begin{align*}
R_N - R_k &\le \log(N/k) \\
\frac{R_N}{R_k} &\le \frac{N}{k} \bigg(1 + \frac{\sigma^2}{P g^2}\bigg),
\end{align*}
with $L=1, a = 0$ and $h_1 = g$. These bounds coincide exactly with the corresponding upper-bounds obtained in \textit{Example~1}. This indicates that the bounds in \eqref{eqn:ecgalDirectAddGap} and \eqref{eqn:ecgalDirectMultGap} are actually tight.

\textit{Remark 2:} The asymptotic expressions of $\prod_{i=1}^l (\beta_i^N)^2$ and $\prod_{i=1}^l (\beta_i^k)^2$ in \eqref{eqn:betaProdPs2infty} and \eqref{eqn:betaProdPs2zero} computed for large $N$ and $k$ result in smaller values of the ratio $\frac{SNR_{t,N}^{opt}}{SNR_{t,k}^{opt}}$ computed in \eqref{eqn:ecgalSNRratioPs2infty} and \eqref{eqn:ecgalSNRratioPs2zero} compared to the corresponding values computed using the expressions for $\beta_i^N$ and $\beta_i^k$ in \eqref{eqn:optBetal4ecgalN} and \eqref{eqn:optBetal4ecgalk}, respectively, for any $N$ and $k$. This implies that the upper bounds on the additive ($\overline{R}_N - \overline{R}_k$) and multiplicative ($\overline{R}_N/\overline{R}_k$) gaps in \eqref{eqn:ecgalDirectAddGap} and \eqref{eqn:ecgalDirectMultGap} computed for asymptotically large $N$ and $k$ are smaller compared to the corresponding bounds $R_N - R_k$ and $R_N/R_k$ obtained for any $N$ and $k$. However, it can be proved that $\overline{R}_N - \overline{R}_k$ and $\overline{R}_N/\overline{R}_k$ approximate $R_N - R_k$ and $R_N/R_k$, respectively, within constants, depending on various system parameters, {\it i.e.} $R_N - R_k \le \overline{R}_N - \overline{R}_k + c$ and $\frac{R_N}{R_k} \le \frac{\overline{R}_N}{\overline{R}_k}(1+d)$, where $c$ and $d$ are two constants dependent on system parameters. For instance, for the network in \textit{Example 2}, we can show that $R_N - R_k \le \overline{R}_N - \overline{R}_k + \frac{h_2^2/h_1^2}{2 \ln 2}$ and $\frac{R_N}{R_k} \le \frac{\overline{R}_N}{\overline{R}_k}(1+\frac{\sigma^2}{P}(h_1^{-2} + h_2^{-2}))$.

\section{Conclusion and Future Work}
\label{sec:conclFW}
Computing the maximum end-to-end rate achievable with analog network coding in general layered relay networks is an important but computationally intractable problem. Previously, this problem is addressed under certain assumptions on input signal factors and received SNRs at the relay nodes. However, such assumptions lead to suboptimal end-to-end performance in general, and in certain communication scenarios it is desirable to compute the optimal rates without such assumptions.

We introduce an approach based on the notion of \textit{network simplification} to approximate the optimal ANC rate within small additive and multiplicative gaps in the Gaussian $N$-relay diamond network and a class of symmetric layered networks while simultaneously reducing the computational complexity of solving this problem. One of our main contributions is a result that states that in a layered network, each layer increases the additive gap between the optimal performance of analog network coding without and with network simplification by $\log(N/k)^2$ and similarly, the multiplicative gap by $(N/k)^2$. To the best of our knowledge, this work provides the first characterization of the performance of network simplification in general layered amplify-and-forward networks. Also, our results suggest a new approach to approximate the optimal ANC rates while allowing for the computation of the additive and multiplicative gaps simultaneously. In future, we plan to extend this work to more general networks and to refine the upper bounds on the additive and multiplicative gaps computed in this paper.

\end{document}